\documentstyle[11pt,aaspp]{article}
\def\ltsima{$\; \buildrel < \over \sim \;$}
\def\simlt{\lower.5ex\hbox{\ltsima}}
\def\gtsima{$\; \buildrel > \over \sim \;$}
\def\simgt{\lower.5ex\hbox{\gtsima}}

\def\etal{{\it et al. }}

\journalid{}{}
\articleid{}{}
\lefthead{Hernquist \& Mihos}
\righthead{}
\slugcomment{Submitted to the {\it Astrophysical Journal}}

\begin{document}

\title{Excitation of Activity in Galaxies by Minor Mergers}

\author{Lars Hernquist\altaffilmark{1} and J. Christopher Mihos}
\affil{Board of Studies in Astronomy and Astrophysics,\break
	 University of California, Santa Cruz, CA 95064\break
         lars@lick.ucsc.edu, hos@lick.ucsc.edu}

\altaffiltext{1}{Alfred P. Sloan Foundation Fellow, Presidential Faculty
Fellow}

\begin{abstract}

Mergers between gas--rich disks and less--massive dwarf galaxies are studied
using numerical simulation.  As the orbit of a satellite decays through
dynamical friction, the primary disk develops large-amplitude spirals in
response to its tidal forcing.  While these features arise in both the stars
and the gas in the disk, the non--axisymmetric structures in the gas differ
slightly from those in the stars.  In particular, as a consequence of the
formation of strong shocks in the gas and the effects of radiative cooling,
the gas response tends to lead the stellar response, enabling the stars to
strongly torque the gas.  These torques deprive the gas of its angular
momentum, forcing a significant fraction of it into the inner regions of the
disk.  Depending on the detailed treatment of the gas physics and the
structure of the primary galaxy, the gas can also condense into dense knots
in reaction to the gravitational perturbation of the dwarf, which later sink
to the center of the primary by dynamical friction against the stellar
background.

The radial inflows induced by these mergers accumulate large quantities of
interstellar gas in the nuclear regions of the host disks.  In some cases,
nearly half of all the gas initially distributed throughout the disk winds
up in a dense ``cloud'' several hundred parsecs in extent.  The models
reported here do not include star formation and, so, we cannot determine
the ultimate fate of the gas.  Nevertheless, given the high densities in
the nuclear gas, it is plausible to identify these concentrations of dense
gas in the remnants with those accompanying intense starbursts in some active
galaxies.  Therefore, the calculations here provide a framework for
interpreting the origin of nuclear activity in otherwise quiescent disk
galaxies.  To the extent that galaxy formation is a chaotic process in which
large structures are built up by the accretion of smaller fragments, our
models may also be relevant to starbursts and the onset of nuclear activity
in proto--galaxies at high redshifts.

\end{abstract}


\vfil\eject

\section{Introduction}

A large fraction of all galaxies are inferred to exhibit unusually high
rates of star formation within their central regions.  The
most spectacular examples of this phenomenon are the so--called
ultraluminous infrared galaxies, discovered originally by the IRAS
observatory ({\it e.g.} Soifer {\it et al.} 1984a,b), some of which
have infrared luminosities well in excess of $10^{12} L_\odot$
(for a review, see Soifer, Houck \& Neugebauer 1987).  While
rare, the ultraluminous IRAS galaxies are now thought to be the
dominant population of extragalactic sources at infrared luminosities
$L_{ir} \simgt 10^{11.3} L_\odot$ and are more common than quasars
having corresponding bolometric luminosities $L_{bol} \simgt 10^{12}
L_\odot$ ({\it e.g.} Soifer {\it et al.} 1987; Soifer {\it et al.}
1989; Scoville \& Soifer 1991; Sanders {\it et al.} 1991; Sanders 1992).

It is generally believed that the luminous infrared galaxies are
powered by unusually high rates of star formation in their nuclei,
although some have spectra reminiscent of active galactic nuclei and
may well contain buried ``monsters.''  (See, {\it e.g.} Sanders \etal
1988ab, Carico {\it et al.} 1990, Condon {\it et al.} 1992.) Supporting
the starburst model are ground--based surveys which indicate that
these objects are typically rich in molecular gas ({\it e.g.} Young
{\it et al.} 1984, 1986; Sanders \& Mirabel 1985; Sanders {\it et al.}
1987) and that more than 50\% of the gas is usually concentrated within
nuclear regions less than one kpc in radius ({\it e.g.} Scoville {\it
et al.} 1986; Sargent {\it et al.} 1987, 1989; Sanders {\it et al.}
1991; Tinney {\it et al.} 1990; Scoville {\it et al.} 1991).  For
example, Arp 220 has an infrared luminosity of $1.5 \times 10^{12}
L_\odot$ and a nuclear concentration of gas a few hundred parsecs in
radius containing of order $2\times 10^{10} M_\odot$ of molecular gas
(see, {\it e.g.} Scoville 1992).

While the origin of these ultraluminous IRAS galaxies is not entirely
clear, a consensus has emerged that these sources are triggered by
collisions involving at least one gas--rich disk galaxy.  This suggestion
appears especially firm for the brightest IRAS galaxies, which invariably
display telltale signs of mergers, such as multiple nuclei, tidal
tails, loops, and shells ({\it e.g.} Allen {\it et al.} 1985; Joseph \&
Wright 1985; Armus {\it et al.} 1987; Sanders {\it et al.} 1988a,b;
Kleinmann {\it et al.} 1988; Sanders 1992).  Self--consistent numerical models
support this notion and demonstrate that the gravitational torques
attending the merger of two gas--rich disks are capable of driving a
large fraction of all the gas to the center of the merger remnant
({\it e.g.} Negroponte \& White 1983; Noguchi 1991; Barnes \& Hernquist 1991),
which may fuel intense star formation (Mihos {\it et al.} 1991, 1992;
Mihos 1992; Mihos \& Hernquist 1994a,b).  The amounts of gas involved,
the final distribution of this gas in the remnant nucleus, and the
time--scales over which it is driven to the center of the remnant are
all in accord with constraints derived from observed properties of
infrared--luminous galaxies.

While ``major'' mergers of comparable--mass spirals are the most
striking example of galaxy collisions, they are less common than ``minor''
mergers of satellites or dwarfs with larger galaxies.  Indeed, simple
estimates employing Chandrasekhar's (1943) description of dynamical
friction show that the {\it typical} galaxy, irrespective of type,
has probably accreted at least several 10's of percents
of its mass in the form of discrete subunits (Ostriker \& Tremaine
1975; Tremaine 1981).  This is to be contrasted with the situation for
major mergers, which leave remnants structurally akin to ellipticals
({\it e.g.} Barnes 1988, 1992), implying that at {\it most} $20\%$ of
all galaxies have experienced a bona fide major merger event, assuming
that all present--day ellipticals formed in this manner.  While it is
beyond the scope of the present discussion to debate the merits and
faults of the ``merger hypothesis'' (Toomre \& Toomre 1972; Toomre
1977), it seems likely that not all ellipticals originated in major
mergers.  If so, then minor mergers are at least an order of magnitude
more common than major mergers, averaged over the age of the universe.
It should be noted, however, that these estimates are subject to
considerable uncertainty (see, {\it e.g.} Hernquist 1991a, 1993a);
nevertheless, they admit the possibility that minor mergers have
played a dominant role in shaping {\it all} galaxy types, a
viewpoint supported by models in which galaxy formation is not a
smooth process ({\it e.g.} Searle \& Zinn 1978; Dubinski \& Carlberg
1991; Katz \& Gunn 1991; Katz 1991, 1992).

N--body simulations demonstrate that minor mergers can drive
structural evolution in disks without completely destroying them
(Quinn \& Goodman 1986; Quinn {\it et al.} 1993).  Among the
properties of disk galaxies which have been attributed to minor
mergers are warps in stellar disks ({\it e.g.} Quinn {\it et al.}
1993), kinematic anomalies in galaxy halos ({\it e.g.} Larson 1990),
dynamical heating of disks (Toth \& Ostriker 1992) and the origin of
``thick'' disks ({\it e.g.} Carney {\it et al.} 1989; Quinn {\it et
al.} 1993; Walker {\it et al.} 1994a), structural peculiarities of some
bulges ({\it e.g.} Hernquist \& Quinn 1989), and the origin of
``amorphous'' disk galaxies ({\it e.g.} Hernquist 1989).  (For a recent review,
see Majewski 1993.)

In fact, the nature of galactic disks following accretion events
has prompted the suggestion ({\it e.g.} Schweizer 1986, 1990) that
minor mergers may contribute to the formation of S0 galaxies,
particularly those having ``X-structures'' (Walker {\it et al.}
1994b).  However, aside from issues such as spheroid building and
disk survival, minor mergers must also explain differences
in the gas content between late-type disk galaxies and S0's in
this scenario.  While S0 disks contain widely varying amounts of gas,
they are typically gas--poor in relation to late--type disk galaxies
(Thronson \etal 1989; Bregman \etal 1990), and display little or no spiral
structure.  Therefore, an understanding of the evolution and final fate
of the ISM in satellite mergers is needed to judge this ``minor merger
hypothesis'' for building S0 disks.

If many occurrences of starbursts in galaxies are triggered by mergers
between galaxies, as suggested by the IRAS discoveries, then
it is natural to inquire whether or not satellite accretions
could produce similar effects.  Hernquist (1989, 1991b) examined this
possibility using a hybrid N--body/hydrodynamics code to follow the
response of disks containing both stars and gas to the tidal forcing of
cannibalized satellites.  His models demonstrate that the two
components react differently to the gravitational torques
induced by the dwarf, ultimately driving large amounts of gas to the
center of the disk.  These findings are qualitatively similar to those
obtained later by Barnes \& Hernquist (1991), and also those of Noguchi
(1987, 1988, 1991) and Noguchi \& Ishibashi (1986) who studied
transient encounters between self--gravitating disks and external
perturbers represented as rigid potentials.  However, Hernquist (1989)
considered only a limited range of parameters and was not able
to determine if the effect is generic to accretion events.

In this paper, we extend the computations of Hernquist (1989, 1991b)
by employing more physical models and by considering a range of
structural parameters for the merging galaxies.  Our results
indicate that while the tendency of tidal torques to promote
radial gas inflows during accretion events does not depend
sensitively on the microphysics used in the simulations,
the structure of the primary galaxy is critical and the strength of
the inflow and ensuing starburst are affected by, {\it e.g.},
whether or not the primary contains a dense, central bulge
(Mihos \& Hernquist 1994b,d).  Unfortunately, the computational
expense of calculations like those reported here makes it
impractical to study a large region of parameter space.
Nevertheless, our results reinforce the notion that galaxy
collisions, in particular minor mergers, play an important
role in triggering unusual activity in galaxies, but that this
role may vary among galaxy types and may have been particularly
significant at earlier phases of galaxy evolution, assuming that disks
formed before most bulges.

\section{Methods}

\subsection{Evolution Code}

The simulations described here were performed with a hybrid
N--body/hydro\-dynamics code for evolving composite systems of collisionless
matter and gas (Hernquist \& Katz 1989).  Gravitational forces are computed
with a hierarchical tree algorithm (Barnes \& Hut 1986), optimized for
vector architectures (Hernquist 1987, 1990b), with a tolerance parameter,
$\theta = 0.6$, including terms up to quadrupole order in the multipole
expansions.

Gravitational forces are softened using a cubic spline (Goodman \&
Hernquist 1991), with different types of particles having different
softening lengths (Barnes 1984, 1985a,b).  In the models described below,
disk, halo, and bulge particles have softening lengths $\epsilon _d =
0.08$, $\epsilon _h = 0.37$, and $\epsilon _b = 0.06$, respectively, in
the dimensionless units defined in \S 2.2.  These values were chosen so
that the softening length of each species is nearly equal to the
mean-interparticle separation at the half-mass radius of each
component.  We tested the integrity of this choice by performing
simulations of isolated disks in which a softening length
$\epsilon = 0.08$ was used for all components, which enabled us
to select an appropriate timestep for experiments in which
different softening lengths were used for the various species.
{}From these test calculations we found that employing slightly larger values
of $\epsilon$ for the coarsely-resolved components (the disk and halo)
mitigates two-body effects relative to simulations in which all particles
interact via the softening length assigned to the most finely resolved
component.

A Lagrangian technique known as smoothed particle hydrodynamics (SPH;
{\it e.g.} Lucy 1977; Gingold \& Monaghan 1977; Monaghan 1985, 1992)
is used to model the disk gas.  In this approach,
gas is partitioned into fluid elements which are represented
computationally as particles.  These particles obey equations of
motion similar to the collisionless particles used to represent the
stars and dark matter, but include terms to describe local effects in
the fluid, such as pressure gradients, viscous forces, and radiative
heating and cooling.  The implementation we use is similar to that
given in Hernquist \& Katz (1989).  In particular, the SPH particles
have their own time--steps, as determined by the Courant condition
(Monaghan 1992; Hernquist \& Katz 1989, eq. 2.37), with Courant number
$C=0.3$.  A conventional form of artificial viscosity is employed
(Hernquist \& Katz 1989; their eqs. 2.22 and 2.23), with the
parameters $\alpha = 0.5$ and $\beta = 1.0$.

As in Hernquist \& Katz (1989) the SPH particles in our models have
their own smoothing lengths, to optimize the range in spatial scales
resolved by the code.  The smoothing lengths are specified by
requiring that each SPH particle have a certain number of neighbors,
${\cal N}_s$, within a volume of radius two smoothing lengths, and
smoothed estimates are symmetrized to preserve momentum conservation
in the manner described by Hernquist \& Katz (1989; their eq. 2.16).

The best choice of ${\cal N}_s$ is somewhat controversial.  A smaller value
implies that smaller scales can be resolved, but at the expense of
introducing larger errors in the local estimates of smoothed
quantities such as the density.  Previous calculations similar to
those discussed here generally employed ${\cal N}_s \approx 30$ ({\it
e.g.} Hernquist 1989, 1991b; Barnes \& Hernquist 1991, 1992a; Hernquist
\& Barnes 1991).  However, empirical and analytic studies of error
propagation in SPH suggest that larger values are preferable for
higher accuracy ({\it e.g.} Rasio \& Shapiro 1991; Rasio 1991; see,
also Hernquist 1993d).  In deference to these analyses, we have chosen
to employ ${\cal N}_s = 96$ in the calculations here, although
it should be noted that our present results are in essential
agreement with the earlier ones using smaller values of ${\cal N}_s$.

Several modifications were incorporated into the original TREESPH code
of Hernquist \& Katz (1989) for our application.  From a comparison of
simulations employing an explicit treatment of radiative heating and
cooling in the gas with others using an isothermal equation of state
(Barnes \& Hernquist 1994), we opted to maintain the gas in the models
here at a fixed temperature.  Owing to limitations imposed by mass
resolution with a finite number of particles, we are unable to
accurately describe the structure of a multi--phase medium, as in more
realistic models of the interstellar medium in galaxies ({\it e.g.}
McKee \& Ostriker 1977).  Hernquist (1989) skirted this issue by
inhibiting radiative cooling below some cut--off temperature, $T_c$,
suppressing the formation of dense clumps of cold gas.  For $T_c \sim
10^4$ K, most of the gas in his models and that in those of, {\it
e.g.}, Barnes \& Hernquist (1994) resides close to this cut--off, owing
to the short radiative cooling time--scales of interstellar gas.  As a
result, simulations with an isothermal equation of state differ little
from those employing a more ``realistic'' treatment of the gas
physics.  The approach chosen here, {\it i.e.} requiring the gas to
remain at a constant temperature, simplifies the interpretation of
the results and enables us to infer the effect of varying a single
parameter on the evolution of the system.  (For a discussion, see
{\it e.g.} Barnes \& Hernquist [1992b, 1993] and for details, see
{\it e.g.} Barnes \& Hernquist [1994] and Mihos \& Hernquist [1994b].)

The equations of motion are integrated using a time--centered
leap--frog algorithm ({\it e.g.} Press {\it et al.} 1986).  The
collisionless particles are integrated on a single, fixed time--step
of $\Delta t = 0.16$ in models without bulges and $\Delta t = 0.08$
when bulges are included (see Hernquist 1992, 1993c).  Scaled to values
appropriate for the Milky Way, these correspond to physical times of
roughly $2\times 10^6$ and $10^6$ years, respectively.
The SPH particles are allowed to have time--steps smaller than this
by factors of powers of two in order to satisfy the Courant condition.
In the most extreme cases we considered, some gas particles had
time--steps as small as $1/128$--th of the large system time--step for
brief intervals during the simulations.  For an isothermal equation of
state, the total system energy is not conserved since it is assumed that heat
generated by adiabatic compression and shocks is immediately lost from
the system.  Even so, the total energy was usually conserved to about
$1\%$ accuracy, and from simulations employing an ideal gas equation
of state with an explicit treatment of radiative heating and cooling,
in which energy losses can be monitored, we infer that integration
errors in our models result in an error in energy conservation at the
level of $0.1\%$.

\subsection{Galaxy Models}

The approach used to construct stable galaxy models consisting of
several distinct, interacting components, is described in
detail by Hernquist (1993b).  Unlike the technique employed by
Hernquist (1989, 1991b), our method permits us to include the
self--gravity of the halos and bulges of the galaxies, rather than
modeling them by rigid potentials.  The
self--gravity of the halos, in particular, contributes significantly
to the torque responsible for the orbital decay of the merging
objects (Walker {\it et al.} 1994a)

As in previous studies which did not include gas (Hernquist 1992,
1993c), the primary galaxies in our simulations include exponential
disks of stars, which initially follow the density profile
$$
     \rho _d (R,z) \, = \, {{M_d}\over{4\pi h^2 z_0}} \, \exp (-R/h)
      \, {\rm sech}^2 \left ( {{z\over{z_0}}} \right ) \, ,
       \eqno(2.1)
$$
where $M_d$ is the disk mass, $h$ is a radial scale--length, and
$z_0$ is a vertical scale--thickness.  These disks are embedded in
self--consistent ``dark''
halos, which have density profiles
$$
     \rho _h (r) \, = \, {{M_h}\over{2\pi ^{3/2}}} {{\alpha}\over
      {r_c}} \, {{\exp (-r^2/{r_c^2})}\over{r^2+\gamma^2}} \, ,
       \eqno(2.2)
$$
where $M_h$ is the halo mass, $r_c$ serves as a cut--off radius,
$\gamma$ is a ``core'' radius, and $\alpha$ is a normalization
constant defined by
$$
     \alpha \, = \, \left [1 \, - \, \sqrt{\pi} q \exp (q^2)
      \left ( 1 - {\rm erf} (q) \right ) \right ] ^{-1} \, ,
       \eqno(2.3)
$$
where $q=\gamma / r_c$.  The abrupt truncation of the
halos as $r \rightarrow r_c$ is not intended to be a realistic
description of the matter distribution in actual galaxies, but
enables us to construct objects having plausible rotation curves
in regions dominated by the luminous material, without requiring
the investment of large amounts of computer resources to follow
the trajectories of loosely bound particles at large radii.
The particles used to represent the disk gas follow roughly
the same density profiles as the disk stars, equation (2.1), but their
vertical scale--height depends on the temperature of the gas
and $R$.  For a gas temperature $T=10^4$ K, the gas layer is
thinner than the stars for vertical stellar dispersions similar to the
Milky Way.

In some cases, we include optional bulges in the primary
galaxies, using an oblate form of the potential--density pair
proposed by Hernquist (1990a) for spherical galaxies and bulges:
$$
     \rho _b (m) = {{M_b}\over{2\pi a c^2}} {1\over{m(1+m)^3}}
      \eqno(2.4)
$$
where $M_b$ is the bulge mass, $a$ is a scale--length along the major
axis, $c$ is a scale--length along the minor axis, and
$$
     m^2 = {{x^2 +y^2}\over{a^2}} + {{z^2}\over{c^2}} \cdot
      \eqno(2.5)
$$
(See, for example, Dubinski \& Carlberg 1991.)

Particles are distributed in space according to the density profiles
specifying the structure of each component included in a particular
model.  Particle velocities are initialized using moments of the
Vlasov equation and by approximating the velocity distributions by
Gaussians (Hernquist 1993b).

In what follows, we express physical quantities in
dimensionless units in which the gravitational constant,
$G=1$, the disk exponential scale--length, $h=1$, and the disk mass,
$M_d =1$.  If we relate these to the properties of the Milky
Way suggested by photometric and kinematic studies ({\it e.g.} Bahcall
\&  Soneira 1980; Caldwell \& Ostriker 1981), {\it i.e.} $h=3.5$ kpc
and $M_d = 5.6 \times 10^{10} M_\odot$, then unit time and velocity
are $1.31 \times 10^7$ years and $262$ km/sec, respectively.

The results presented in \S 3, which describe in detail a minor merger
between a gas--rich disk and a smaller companion, employed a galaxy
model not having a bulge and whose stellar disk and halo parameters
are $z_0 = 0.2$, $M_h = 5.8$, $\gamma = 1$, and $r_c = 10$.  The
Toomre $Q$ parameter (Toomre 1963) varies weakly with radius in these
models; we normalize it so that for the stars $Q=1.5$ at a
solar radius, $R_\odot = 8.5/3.5$.  In this model, the gas comprises
$10\%$ the total mass of the disk; {\it i.e.} the stellar disk has a
mass of $0.9$ and the disk gas has a mass of $0.1$ in our
dimensionless system of units.  For a temperature of $10^4$ K, the gas
particles are distributed initially with the density profile of
equation (2.1) but with a vertical thickness depending on $R$.  If an
isothermal equation of state is adopted, the vertical width of the gas
layer increases with distance from the center of the galaxy.  This
effect is included in an approximate manner when the disk is created
and small departures from the actual vertical profile mostly dissipate
on a vertical sound--crossing time once the galaxy is allowed to
evolve.

Figure 1 shows the result of a time--integration of one of these
galaxy models in isolation.  In this example, the luminous matter
comprises stellar and gas disks of mass $0.9$ and $0.1$, respectively.
A total of 32,768 particles represent each of the two collisionless
components, {\it i.e.} the disk and the halo, and 16,384 SPH particles
describe the interstellar gas.  In this case, the gas was taken to be
isothermal with a temperature $T=10^4$ K.  As is evident from Figure 1, the
disk rapidly develops a spiral pattern that persists throughout the
time period covered by the simulation.  This structure probably results
from the swing-amplification of noise in the potential by the disk (Toomre
1981, 1990), although there are indications that this explanation may not be
complete ({\it e.g.} Sellwood 1989a).  We are encouraged that our
models are well-behaved over the timescales needed to study satellite
accretions like those discussed below.  In particular, the disks
are stable and do not develop spontaneous inflows of gas.  Over much
longer timescales, however, the models are bar-unstable owing to
discreteness noise in the halo potential which can be amplified by
the disk (Sellwood 1989b).  This effect is negligible for the simulations
described in \S\S 3 and 4, but complicates efforts to investigate
orbital decay from polar and retrograde orbits (Walker {\it et al.}
1994a).

\subsection{Satellite and Encounter Parameters}

The satellite galaxies in our simulations are represented by spherical
versions of the model given by equations (2.4) and (2.5).  Our fiducial
simulation, described in \S 3, employed a satellite with mass
$M_{sat} = 0.1$ and scale-length $a=0.15$, so that its mean half-mass
density is roughly the same as the mean density of disk material
in the same volume near the center of the primary.  This choice
ensures that some portion of the satellite will survive to the
center of the disk in cases where orbital decay occurs rapidly.
The mass distribution of the satellite is truncated so that it is initially
within its tidal radius.  We also ran models with either more massive or
denser satellites; the results of these calculations are discussed in \S 4.

Initially, the dwarfs are placed on circular orbits at 6 scale-lengths
from the center of the primary.  We ran smaller calculations varying
the size of the orbit and the results were qualitatively similar to
those described here, but required longer integrations in cases where
the satellite's orbit was more extended.  We also varied the inclination of
the orbit plane.  In the models described in detail below, the orbit
plane is initially inclined relative to the disk by 30 degrees, and the
satellite orbits in a prograde manner.  Polar and retrograde orbits require
prohibitive amounts of cpu time for the mass ratios we consider (see Walker
{\it et al.} 1994a).  Less-inclined orbits decay slightly more rapidly;
the qualitative outcome is similar to mildly inclined orbits.

Clearly, the assumption of a circular orbit is not entirely realistic,
given the eccentricities of the Milky Way's companions.  Our choice
is motivated by practical considerations which force us to consider
mergers which occur rapidly.  More sophisticated approaches for
computing the self-gravity of the interaction and the use of parallel
hardware may make it possible to examine a wider set of parameters
in the near future.

\section{Galaxy Models}

In this section we describe the outcome of a merger between
a primary galaxy consisting of a disk and dark halo and a satellite
companion.  The parameters were chosen to resemble the encounter
reported by Hernquist (1989), but here we include the full
self-gravity of each component.

\subsection{Overview of Merger Dynamics}

Figure 2 presents a set of face-on images of the merger.  Owing
to dynamical friction against the disk and the halo, the orbit
decays rapidly and the satellite sinks to the center of the disk within only
a few orbital periods; the entire time-period displayed in
Figure 2 spans roughly $1.2 \times 10^9$ years.  In response to
tidal torques from the dwarf, the primary disk exhibits
large-amplitude non-axisymmetries that are particularly striking
at intermediate times between $t \approx 30$ and $t \approx 70$.
At no time does the disk develop an extended bar, although the
structures are linear in appearance near the center.  As usual,
the regions leading the satellite are mostly devoid of material
because of the gravitational perturbation exerted on stars there
by the dwarf (Quinn \& Goodman 1986).  Once the merger is
complete, the disk supports some weak spiral structure that
eventually mixes away, in contrast to the longer-lived features
present in isolated disks, like those shown in Figure 1.

Figure 3 shows an edge-on view of the encounter.  As has been noted
previously (Quinn \& Goodman 1986; Quinn {\it et al.} 1993),
the satellite sinks first into the disk plane before
plunging into the center of the primary,
warping the outer parts of the disk.  Depending on the shape of
the halo, these warps can persist for a significant fraction of
a Hubble time before they are washed out by phase-mixing
(Quinn {\it et al.} 1993; Dubinski 1994).  The inner parts of the disk are
heated vertically, increasing its thickness and the vertical
dispersion of stars there.  The characteristics of post-merger
stellar disks like that in Figure 3 are compared with observed
properties of thick galactic disks by Quinn {\it et al.} (1993)
and Walker {\it et al.} (1994a) who argue that accretion events
are a viable mechanism for producing these features.

Viewed from afar, the gas in the disk behaves qualitatively like
the stars do, as can be seen in Figure 4.  The spirals which
develop in the gas are crisper than their stellar analogues,
however.  This effect, together with the time-dependent nature
of the tidal field, allows the stellar response to torque the
gas, forcing much of it into the center of the disk.

Material stripped from the satellite during its plunge through the
disk forms both leading and trailing ``streamers'' which eventually
wind up to resemble a spiral feature, as can be seen in Figure 5.
This pattern is kinematic, transient in nature, and fades
completely in a few orbital periods beyond the last frame shown in
Figure 5.  For the choice of parameters in this model, roughly
$40\%$ of the satellite material sinks to within one scale-length
of the disk center.  At late stages of the encounter, however, a
dense blob of gas accumulates in the middle of the disk which
completely disrupts the dwarf during the final stages of the
merger, in a manner analogous to that in the model of Hernquist
(1989).

The final stages of the merger are illustrated in Figure 6, which
shown the gas in the inner parts of the disk, magnified ten times
relative to the images in Figure 4.  As the dwarf approaches the
central regions of the disk, the non-axisymmetric response
dominates both the gas and stellar distributions.  Near the
center, this response takes the form of a linear feature which
resembles a bar and which is clearly visible by $t=53.6$ in
Figure 6.  As the dwarf plunges into the center, the gas in
this feature is forced onto self-intersecting orbits which
dissipate its kinetic energy by forming shocks.  The internal
energy added to the gas is radiated away immediately, with
our assumption of an isothermal equation of state.  (The
results are not changed qualitatively if an ideal equation
of state is used together with a more physical description
of radiative cooling, owing to the short cooling times of the
gas in this region [Barnes \& Hernquist 1994].)  Because of
this rapid dissipation of energy the response of the gas differs
from that of the stars, which are not required to follow closed
orbits.  As we show in \S 3.2, this difference is such that the
stars torque the gas, removing its angular momentum, and allowing
it to flow rapidly into the center of the primary.

Figure 7 shows mass profiles of the gas at various stages during the
merger.  By the time of the final panel in Figure 6 the collapse of
gas into the center is essentially complete.  As can be inferred from
Figure 7, by the end of the sequence shown in Figure 6, approximately
$45\%$ of the gas initially spread throughout the disk has collapsed
into a region only a few hundred pc across, when scaled to the values
given in \S 2.2.  The detailed structure of this nuclear accumulation
of gas is not accurately modeled by our simulations because of
gravitational softening.  Nevertheless, the gas densities in this
region are sufficiently high, in principle, to trigger intense
starburst activity (Mihos \& Hernquist 1994b).

The onset of the catastrophe shown in Figures 6 and 7 is accompanied
by the tendency of cool gas in our models to fragment on small scales
under its own self-gravity in regions containing self-intersecting
flows.  Evidence for this effect is clearly visible at intermediate
times in Figure 6, between $t=53.6$ and $t=60.0$.  However, as
discussed further in \S 4, the radial inflows are virtually unaffected
if the gas temperature is raised to inhibit fragmentation or if the
self-gravity of the gas is explicitly ignored, essentially preventing
the gas from breaking apart into clumps.  These findings support the
view of Noguchi (1988) who argued that the self-gravity of the gas
plays an unimportant role in the onset of tidally induced nuclear gas
flows in disks.  In \S\S 3.2 and 4.1, we discuss physical mechanisms
responsible for the outcome shown in Figures 6 and 7.

\subsection{Angular Momentum Transport \& Torques}

The physical nature of the nuclear gas inflow in our fiducial
model can partly be understood by examing the evolution of the
angular momentum of the gas driven to the center of primary
during the merger (Barnes \& Hernquist 1994).  In Figure 8, we
show the orbital, spin, and total angular momentum of the halo
(panel a), the disk stars (panel b), the gas which accumulates
in the nucleus (panel c), and the portion of the satellite which
sinks to within one scale-length of the center (panel d).  The
quantities shown were computed in the manner described by Hernquist
(1992).

As in all such events, the orbital angular momentum of the merging galaxies
is efficiently converted into spin angular momentum of the various
components through tidal torquing.  From panel (d) of Figure 8, it can
be seen that the orbital angular momentum of the dwarf decays
monotonically.  In response, the halo and stellar disk are spun up
(as seen in Figures 8(a) and 8(b)), accepting comparable amounts of
the satellite's orbital angular momentum.  Following its disruption by the
gas cloud at the center of the disk, the satellite debris acquire a small
amount of spin angular momentum as its remains are incorporated into a
diffuse ring orbiting the nucleus (see, {\it e.g.} Figure 3 of
Hernquist 1989).

The angular momentum of the gas deposited into the center of the
disk declines monotonically until $t\approx 80$, after which it
remains nearly constant as the inner regions of the disk settle
into a new equilibrium.  As can be seen from panel (c) of Figure
8, the angular momentum of this gas has declined by nearly two
orders of magnitude during the merger, enabling it to accumulate
on the scales shown in Figure 7.  This loss of angular momentum is
most rapid during the stages of the encounter shown in Figure 6.

To interpret the loss of angular momentum by the central gas
accumulation, we identified the particles in this object when the
merger was complete and computed the various torques on this
group of particles during the course of the merger.  Figure 9
shows the time evolution of the torque on this gas.  In panel (a),
the magnitude of the full torque on the gas is compared with the
rate of change of its internal (spin) angular momentum.  The two
curves track each other quite well, proving that the gas loses
angular momentum in response to its physical torquing rather than,
for example, as the result of numerical errors.  Small differences
between the solid and dotted curves in panel (a) are consistent with
inaccuracies expected in estimating $|ds/dt|$ from data that are
somewhat coarsely sampled in time.

Panel (b) of Figure 9 compares the gravitational torque acting
on the gas, indicated by the dashed curve, with the hydrodynamical
torque, shown as the dotted curve.  Clearly, the hydrodynamical
contribution to the torque is negligible compared with the
gravitational torque throughout the merger, in agreement with
similar findings by Barnes \& Hernquist (1991) and Barnes \&
Hernquist (1994) for gas flows induced by major mergers.  This
does not mean that hydrodynamical effects are unimportant to the
dynamics, however.  As we emphasize in \S 5, shock heating and
radiative cooling are responsible for the differences in the
dynamics of the gas relative to that of the disk stars, which is
precisely why gas is driven to the center of the remnant.  This
point is made plausible by panel (c) of Figure 9 which shows
the gravitational torque on the gas, broken down by mass component.
The torque is always dominated by its contribution from the disk
stars, rather than that directly from the satellite.  In fact, since
in the final stages of the accretion the orbital phase of the dwarf
tends to lead the gas response,
the direct effect of the satellite on the gas is to torque it up,
averaged over time.  The halo also plays a mostly negligible
role in removing angular momentum from the gas.  Only during the
final stages of the encounter is its influence at all significant,
and even then its contribution to the full torque amounts to less
than $25\%$.

Not surprisingly, the gravitational torque on the central gas from
the disk stars is dominated by material in the inner portions of the
disk.  This is illustrated in panel (d) of Figure 9, where the
gravitational torque from the disk is broken down into its
contributions from stars in quartiles of the initial mass
distribution.  The torque from the innermost two quartiles is
clearly much greater than that from outlying regions and, so,
it is the matter within the initial half-mass radius of the disk
that is mostly responsible for torquing the gas and removing
its angular momentum.

It is also of interest to determine the dominant azimuthal
wavenumber of the disk response that determines the torque on
the gas driven to the center of the remnant.  Figure 10
shows the power in the lowest non-axisymmetric modes in the
inner half mass of the stellar disk as a
function of time during the merger.  The structure is always
dominated by an $m=2$ response.  Higher order modes are nearly
always unimportant, except at late times when a three-armed
spiral ($m=3$) evidently plays a small, but not entirely
negligible role.  It is reassuring, but not unexpected, that
the amplitude of the non-axisymmetric disk response peaks at
times when the torque on the gas is a maximum, as can be seen
by comparing Figures 9 and 10.  The exact nature of the $m=2$
disk response is a bit difficult to characterize.  Throughout
most of the encounter, the disk is highly distorted and sports
a pair of well-defined spirals.  At late stages, the response
near the center is linear in appearance, although this
``bar-like'' distortion is not reminiscent of bars in actual
spirals owing to its short lifetime and extent.

\section{Other Models}

In spite of some differences in the details of the simulations,
the results of our fiducial model are in good agreement with the
findings of Hernquist (1989), as discussed in \S 5.  In his
earlier study, however, Hernquist made no attempt to vary
numerical parameters or those defining the structures of the
merging galaxies.  Below, we describe our attempts to explore
interesting regions of parameter space, based on 20 or so additional
models.

\subsection{Variation in Gas Temperature}

For simplicity, we chose to employ an isothermal equation of
state in the calculations reported here.  Requiring the gas to
remain at a fixed temperature is equivalent to demanding that
it radiatively heat and cool at well-defined rates.  A higher
gas temperature thus corresponds to slightly less efficient
cooling.

We investigated the sensitivity of the results described above to
the temperature of the gas by running four calculations identical in
all respects to our fiducial model, but employing gas temperatures
$T=3\times 10^3, 3\times 10^4, 10^5,$ and $3\times 10^5$ K.
Qualitatively, the outcomes were similar to our fiducial model,
except for the simulation with a $T=3\times 10^5$ K gas, which
exhibited substantially less gas inflow than the others.  At
intermediate stages of the merger, however, the detailed structure
of the gas in the disk depends sensitively on its temperature.
This effect is illustrated in Figures 11 and 12 which are identical
to Figure 6, but for the models with $T=3\times 10^3$ and $T=10^5$
K gases, respectively.

The cooler gas has a greater tendency to fragment under its own
self-gravity in regions supporting flows that self-intersect.  This
trend is obvious in Figure 11, where a number of distinct fragments
form in the gas by $t\approx 57.6$ which eventually sink to the
center of the disk and coalesce. In the evolution summarized in Figure
11, several of the fragments are sufficiently massive that the dynamical
friction they experience is adequate to drive them to the disk center.
We emphasize that this effect is driven by the response of the disk
to the tidal forcing of the dwarf and does not occur in our isolated
models over the temperature range we examined, as we verified by running
a counterpart to the simulation in Figure 1 with a $3000$ K gas.

For higher gas temperatures, the fragmentation process just described
is inhibited, as can be inferred from Figure 12.  The appearance here
is much smoother than that in Figure 6 or especially that in Figure 11.
Also, as suggested by the frame at $t=81.6$, the physical mechanism
which enables the disk stars to torque the gas is more evident.  During
the final stages of the merger, the potential near the center of the
disk changes rapidly owing to the contribution from the satellite,
whose orbit decays rapidly at this late time.  The disk stars do not
dissipate energy and so are less restricted in their choice of orbits
than the gas, which must reside on closed orbits if it is not to
dissipate energy.  In a time-stationary potential that has a large
non-axisymmetric component, gas can find closed orbits without
dissipating a large fraction of its kinetic energy.  In a potential
which varies rapidly with time, as in Figure 12, however, the gas
dissipates energy continuously since the family of closed orbits
changes on the orbital decay timescale of the satellite.  This
results in significant shocking of the gas, as can be seen in
Figure 12 at $t=81.6$, and subsequent loss of energy by the gas
as it instantaneously radiates away its excess internal energy,
as required by the condition that it remain isothermal.

It is interesting that in spite of the differences in detail between
Figures 6, 11, and 12 the amounts of gas driven to the center of the
remnant and the final distributions of this gas are quite similar.
This is illustrated in Figure 13 which shows the final mass profiles
of disk gas in the five simulations in which we varied the gas
temperature.  The quantity of gas driven to the center of the disk
in the simulation employing a $T=10^5$ K gas is only slightly less
than in the runs with lower temperature gases.  Evidently, however,
the catastrophe can be inhibited if the gas is sufficiently hot, as
can be seen from the curve in Figure 13 for the run with a $T=
3\times 10^5$.  We believe that this particular choice of equation
of state is extreme, as implied by estimates of the temperature
of diffuse gas in our own galaxy and the velocity dispersion of
clouds.

There is a weak tendency for the warmer gas to form more diffuse
accumulations in the centers of the disks, as can be inferred from
Figure 14, where we plot the distributions of gas densities
measured from the SPH particles in the various runs following the
merger.  Although the amounts of high density gas do not change
significantly with increasing gas temperature, aside from the
$T=3\times 10^5$ K case, the maximum gas density declines slowly but
monotonically with increasing temperature.  This results from the
larger thermal support of the nuclear gas in the higher temperature
models.  Similar calculations employing an algorithm for converting
the gas into stars (Mihos \& Hernquist 1994d) imply, however, that
this effect is not sufficient to prevent a strong starburst from
occurring in the gas in all cases we have examined (again, aside
from the model with $T=3\times 10^5$ K).

On the basis of Figures 13 and 14, we conclude that nuclear
inflows of gas in disks accreting small satellites are relatively
insensitive to details in the equation of state, provided that
the gas can globally dissipate energy at roughly the same rate as
that in the diffuse component of the ISM in our galaxy or as that
in an ensemble of colliding clouds which maintain a velocity
dispersion similar to the sound speed of a $10^4$ K gas.  This
latter claim is supported by the qualitatively similar nature of
our results to those which have been obtained modeling the gas as
a collection of discrete clouds ({\it e.g.} Negroponte \& White 1983;
Noguchi 1987, 1988, 1991).

\subsection{Variation in Galaxy Structure}

Of more immediate significance are the effects of varying the
structure of the primary galaxy.  To study this, we ran a set of
models, identical in all respects to that described in \S 3, but
including compact bulges in the primary.  Here, we report on one
of the simulations in detail.  The bulge followed the density
profile of equation (2.5) with $a=c=0.2$ and $M_b=1/3$, making the
primary similar to those employed in the stellar-dynamical models
of Hernquist (1993c).  Unlike the rotation curve of the disk in
our fiducial model, which rises slowly over the inner two
scale-lengths of the disk (Hernquist 1992), the rotation curve
of the model with this bugle climbs abruptly to within $\approx 75\%$
of its maximum and then to its peak over roughly two scale-lengths
(Hernquist 1993c), in a manner similar to the rotation curves
of typical early-type spirals.

Figure 15 shows the response of the stellar disk to the sinking
satellite during the merger and Figure 16 shows the response of
the gas in the disk at the same times.  Qualitatively, at this
scale the merger appears similar to that shown in Figures 2 and 3.
A frame-by-frame comparison shows, however, that the tidally
induced structure in this new disk winds up more rapidly than that
in our fiducial model.  There is also a suggestion from the morphology
that the non-axisymmetric features are slightly less well-defined
than those in Figure 2.  This is supported by a modal decomposition
of the stellar disk, shown in Figure 17.  A comparison with
Figure 10 shows that while the non-axisymmetric response is again
dominated by two-armed structures, the maximum power in the $m=2$
mode is suppressed relative to that in the fiducial model,
and this feature lives for a shorter period of time.  In addition, the nature
of the response has a slightly different visual appearance.  While the
inner regions of the disk with no bulge develop a linear, bar-like
structure, the $m=2$ mode in the disk in Figure 15 is morphologically
like a wrapped spiral.  Higher order modes are nearly absent in the
new model.

The response of the gas on this scale is likewise similar in appearance
to that in Figure 3; on smaller scales, however, it is dramatically
altered.  A magnified view of the gas in the inner portions of the
disk is shown in Figure 18. Rather than sporting the linear bar-like
feature shown in the fiducial bulgeless merger (Figure 6), the gas in this
merger forms a tightly wrapped spiral, reducing the dissipation and
inflow of gas into the very center of the primary disk.
This is illustrated in Figure 19, where final mass profiles
are shown for the run with the central bulge and our fiducial model
which has no bulge.  We also show one of our other simulations which
included a less-massive bulge with mass $M_b=1/9$.  In this additional
model, the bulge-scale length was chosen to be 0.14, so that the
mean density was similar to that of the more massive bulge.

Figure 19 shows that the nuclear inflow is dramatically altered by
changes in the mass distribution of the primary galaxy.  Some gas
is driven into the inner regions of the galaxies which include
bulges, but both the amount of gas involved and the ultimate
concentration of this gas are greatly reduced compared to our fiducial
model.  Evidently, the degree to which accretion events can induce
activity in gas-rich spirals depends crucially on the structure of
the primary galaxy.

The robustness of the apparent critical importance
of the structure of the primary galaxy was examined by considering both
denser and more massive satellite galaxies.  These alterations did not
promote more intense inflows of gas than those summarized by Figure 18
for models which included bulges.
The presence of a dense central component appears to inhibit strong
radial inflows of gas, and accompanying starbursts (Mihos \& Hernquist
1994d), under rather broad circumstances.  Clearly, if significantly
{\it less} dense satellites were used they would be torn apart by
tidal forces long before reaching the center of the disks and would
therefore be unable to induce gas inflows of the type discussed above.
In this sense, our results are determined by the relative structure
of the primary and secondary galaxies.

\subsection{Miscellaneous}

We also checked the sensitivity of our results to the numerical
parameters, such as the number of particles in the various components
and those determining the magnitude of the artificial viscosity; no
significant differences were found.  Indeed, the results described in
\S 3 are in good qualitative and quantitative agreement with those of
Hernquist (1989), who used a smaller number of particles, a rigid
halo, larger values for the parameters defining the artificial
viscosity, a smaller value for ${\cal N}_s$, and an explicit treatment
of radiative heating and cooling of the gas, rather than an isothermal
equation of state.  Our inability to
detect any dependence on the numerical parameters does not, of course,
prove that we have obtained converged solutions, since we are able to
vary, {\it e.g.} particle numbers by factors of only a few.
Nevertheless, we are encouraged that our results appear to be relatively
insensitive to what might be expected to be the physically most poorly
constrained quantities, such as the artificial viscosity and SPH
smoothing lengths.

To investigate Noguchi's (1988) claim that the self-gravity of the gas
is unimportant in triggering radial inflows of gas in tidally
perturbed disks, we ran several calculations in which the self-gravity
of the gas was ignored.  One such model is reported by Hernquist
(1991b).  Our results confirm Noguchi's suspicion to the extent that
the onset of the catastrophe is relatively insensitive to this aspect
of the dynamics.  However, the ultimate distribution of the gas driven
to the center of the disk, and hence the density distribution of the
gas, are determined at least in part by the self-gravity of the gas.
For example, in the simulation reported by Hernquist (1991b) roughly
$50\%$ of the gas in the disk was driven to a region $\approx$ 600
pc across.  The amount of gas involved is slightly greater than in
the experiment reported in \S 3.1, since a more massive stellar disk
was employed to maintain the same overall rotation curve for the
galaxy, resulting in more efficient gravitational torquing of the
gas by the stars.  However, the resulting central gas concentration
was more diffuse than in the experiment described in \S 3.1.  In fact,
this difference is hardly surprising, since the gas is essentially
fully self-gravitating once significant amounts of it accumulate in the
central portions of the primary galaxy.

We also varied the fraction of gas in the primary disk.  The results
described in \S\S 3 and 4 were not qualitatively altered by these
changes.  Unfortunately, it is difficult to employ significantly larger
gas fractions than those adopted here, because of the tendency of the
gas to fragment when feedback from star formation is ignored.  We
expect that sufficiently detailed models would confirm our belief
that disks with larger fractions of gas would be even more susceptible
to tidally induced inflows than the simulations reported here.

Finally, we attempted to determine in detail why the presence of a
dense bulge in the primary galaxy suppressed the gas inflow seen in
our fiducial model, by using primaries with no bulges but with more
compact halos, and hence more sharply rising rotation curves than
in our canonical bulgeless galaxies.  Unfortunately, the results were
ambiguous.  The collapse of gas into the center of the primary was
similar to that seen in our fiducial model.  However, this may have
simply reflected the fact that we necessarily used a less massive
halo than in the model describe in \S 3 to mimic the rotation curve
of the model discussed in \S 4.2 which had a bulge with mass
$M_b = 1/3$.  In truth, our simulations are not well-suited for
isolating the detailed dynamics responsible for the outcomes seen
because of the large number of free parameters defining each
calculation.

\section{Discussion}

\subsection{Comparison with Previous Work}

The models analyzed and presented here are most closely related to
those of Hernquist (1989), who also considered the accretion of
less-massive companions by gas-rich disks.  Although our new
calculations differ considerably in detail from those of Hernquist
(1989), our overall conclusions are similar to those expressed in
this earlier study.  In particular, we have reaffirmed the possibility
that minor mergers can induce radial inflows of gas in galactic
disks.  In the case of our fiducial model, nearly $45\%$ of all the
gas initially spread throughout the disk collapses into a region
several hundred parsecs across on a timescale of order a few
hundred million years.  As we show in a companion study (Mihos \&
Hernquist 1994c,d), if a plausible description of star formation is
incorporated into the models, the gas densities in this region are
sufficient to trigger an intense starburst lasting for $\sim 10^8$
years.  As noted by Hernquist (1989), in principle this result
eliminates worries over the observed mismatch between burst times
$\sim 10^8$ years and galactic dynamical times $\sim 10^9$ years
(Scoville \& Norman 1988; Norman \& Scoville 1988; Norman 1988).

These results are in good quantitative agreement with those of
Hernquist (1989), in spite of the differences between the two
computational approaches, as indicated in \S 4.3 above.  One
discrepancy is that our fiducial model suffers a slightly more
intense radial gas inflow than the simulation described by
Hernquist (1989), in which $35\%$ of the disk gas was driven
to the center of the primary.  This difference can be attributed
to Hernquist's use of a more diffuse satellite companion which
suffered correspondingly greater tidal stripping than the
satellite used in our calculations.  Thus, although we employ the same
satellite-to-disk mass as Hernquist (1989), a slightly larger
fraction of the companion's mass survived intact, inducing a
somewhat greater tidal response in the disk.  Clearly, however,
this difference is not relevant to the basic findings of either
study, as a result in better agreement with Hernquist (1989) would
presumably be obtained with our current models if a slightly
less massive satellite were employed.  (For a discussion, see
Walker, Mihos \& Hernquist 1994.)

\subsection{Dynamics of the Response}

Our present investigation goes well beyond those of Hernquist (1989)
in the sense that we have considered the sensitivity of the effects
to some variations in physical and numerical parameters.  For example,
it appears that the onset of gas inflows is relatively insensitive to
the temperature of the gas, provided that it is not extraordinarily
high.  The results are also not affected by feedback from the ensuing
starburst (Mihos \& Hernquist 1994c,d), at least when relatively simple
prescriptions are used to include star formation in the gas (Mihos
\& Hernquist 1994e).

Using our fiducial model, we have established that the gas inflows are
driven by gravitational torquing of the gas by the stellar disk, in
response to the tidal forcing of the companion galaxy, in a manner
noted by other workers in related studies ({\it e.g.} Combes, Dupraz
\& Gerin 1990; Barnes \& Hernquist 1991; Barnes \& Hernquist 1994).
This outcome can plausibly be blamed on the different dynamics
of the gas relative to the stars in response to the time-varying,
non-axisymmetric potential which develops during the final stages
of the merger.  The disk gas will dissipate energy in shocks and
radiate away excess internal energy unless it follows closed orbits;
a requirement not shared by the stellar disk.  As the potential
changes most rapidly, it becomes increasingly difficult for the
gas to avoid rapidly dissipating energy in strong shocks.  The
establishment of these shocks is such that the gas is torqued by
the disk stars (Barnes \& Hernquist 1991), removing angular momentum
from the gas and driving it to the center of the disk.  Regrettably,
the uncontrolled nature of our experiments makes them poorly suited
for examining the detailed orbital response of the various components
at late times in the merger when conditions change most rapidly.
Thus, while we believe that numerical effects are not responsible for
the dynamics seen in the models, we have not demonstrated that,
{\it e.g.} the artificial viscosity plays a negligible role in
the nuclear inflows in our models.  (Although we have performed
simulations in which we varied the magnitude of the artificial
viscosity by factors of several in each direction with inconsequential
results.)  It appears to us that a more fruitful approach to
fully isolate all the dynamics would be to employ more controlled
experiments in which disks are subjected to external, time-varying
perturbations that can be applied systematically.

\subsection{Structure of the Primary Galaxy}

As evidenced by the simulations described in \S 4.2, the structure
of the primary galaxy is critical in determining the nature of gas
inflows that can be induced by accreted companions.  A related
discovery was made by Mihos \& Hernquist (1994a), who found that
the presence of compact bulges in galaxies involved in a major
merger delayed the onset of gas inflows until late in the encounter,
resulting in stronger and more abrupt starbursts than in collisions
between pure disk-halo galaxies ({\it e.g.} Mihos {\it et al.} 1991,
1992).  Our results imply that radial inflows in minor mergers
can not only be delayed but be entirely suppressed by variations in
the structure of the primary.  At present, with our limited set of
simulations, we are unable to isolate the precise mechanism
responsible for this effect, although it may be related to
previous claims that the presence or absence of an inner Lindblad
resonance may play an important role in gas flows near the centers
of disk galaxies ({\it e.g.} Combes 1988; Kenney {\it et al.} 1992).
Our results suggest that extremely violent nuclear starbursts may
be limited to disk galaxies with small bulge to disk ratio. Observationally,
the connection between central starbursts and bulge to disk ratio
is still unclear; however, the wide variation of galaxy properties
along the Hubble sequence (Roberts \& Haynes 1994) would make
any such connection difficult to isolate.

Our results on the sensitivity of the inflow to the presence of bulges
differs from that of Barnes \& Hernquist (1991, 1994), who found that
dissipation in major mergers was not suppressed by the presence of
bulges in the merging galaxies. However, while the bulges used in their
galaxy models were comparable in mass to those employed here, their bulges
were significantly more diffuse.  Therefore, it is not bulge
to disk ratio alone that drives the effects noted here, but rather
the central density of the bulge component. Evidently bulges must have
sufficient density in order to suppress radial inflows of gas during
galaxy mergers.

\subsection{Evolution Along the Hubble Sequence}

By analogy with the merger hypothesis for the origin of elliptical
galaxies (Toomre \& Toomre 1972; Toomre 1977), Schweizer (1990)
has proposed that at least some early-type spirals and S0's might
have resulted from the accretion of less-massive companions by
late-type disk galaxies.  While not designed specifically to
examine this issue, our models generally support this notion.
Although we do not address the ultimate fate of the central
gas clump formed in these encounters, models which include
star formation (Mihos \& Hernquist 1994c) indicate that this
gas is almost completely transformed into stars during an
induced central starburst. Therefore, the remnants of these
encounters may well resemble S0 galaxies, at least to the extent
that they are
much less gas rich than their progenitor disks, contain thickened
disks, and exhibit only rather diffuse spiral structure.  The
latter property is also characteristic of amorphous galaxies ({\it e.g.}
Sandage \& Brucato 1979), suggesting that they may have formed
in this manner.  Perhaps of most relevance are S0
and early-type galaxies possessing central X-structures ({\it e.g.}
Whitmore \& Bell 1988).  In the past, these features have often
been explained by the accretion of material from a passing galaxy
during a transient encounter ({\it e.g.} Hernquist \& Quinn 1989),
although stellar-dynamical models like those reported here have
shown that such features can arise through minor mergers (Walker
{\it et al.} 1994a).  The present simulations demonstrate
further how gas can be removed from the disk in such
minor mergers, leaving behind remnants with diffuse gas content
and morphologies similar to a number of well-studied S0 galaxies
with X-structures (see, {\it e.g.} Walker {\it et al.} 1994b).

\subsection{Relation to Active Galaxies}

Of greatest interest is the possibility that minor mergers may
be responsible for the onset of unusually energetic activity in
the nuclei of some galaxies.  Simulations like those described
in \S\S 3 and 4, but which include a simple prescription for
star formation and feedback, show that mergers like our
fiducial model can produce nuclear starbursts with characteristics
remiscent of those observed in some cases (Mihos \& Hernquist
1994c,d).  Clearly, our calculations are not relevant to those
systems that likely consist of two merging galaxies that are
comparable in mass, as appears to be the case for many of the
ultraluminous sources ({\it e.g.} Sanders 1992).  However, to
the extent that minor mergers are a more common type of event,
the simulations here may be relevant to a larger percentage of
starbursts.  One caveat, of course, is our finding that the
presence of a compact bulge can effectively halt central inflows
of gas induced by accretions, inhibiting nuclear starbursts.  If
this result is supported by future modeling, we must conclude that
starburst activity triggered by minor mergers will be limited to
only certain progenitors; namely, late-type disk galaxies that
are rich in gas.  Such cases are still of great interest, however,
as it may well be the case that disks form before bulges and in
view of the fact that galaxies may generally be assembled in a
clumpy manner ({\it e.g.} Searle \& Zinn 1978).  In that event,
we anticipate that minor mergers may have played a greater role
in triggering nuclear activity in galaxies early in the Universe.

Our models also suggest that any induced central activity may
not occur until the satellite sinks to the central few kiloparsecs
of the primary center. During the early stages of the encounter,
the gas inflow rates are relatively small; rather than driving
central activity, the merger may excite heightened disk star
formation (Larson \& Tinsley 1978; Kennicutt \etal 1987). Models which
include star formation (Mihos \& Hernquist 1994d) indicate that
the star formation rate at these early times remains mostly constant,
increasing only as the nuclear inflow develops.  Accordingly,
studies which seek to link starburst or AGN activity with the
presence of companion galaxies ({\it e.g.} Dahari 1984; Keel {\it et al.}
1985; MacKenty 1989) may actually underestimate the statistical
connection, as galaxy samples selected by the presence of companions
may exclude many late-stage active mergers where the companion is
difficult to discern.

More speculatively, it has been suggested at various times that
galaxy collisions may be related to the onset of various types
of active galactic nuclei (for a review, see Barnes \& Hernquist
1992b).  Owing to the limited resolution of our simulations we
are unable to address this interesting possibility.  Nevertheless,
there is circumstantial evidence which favors this interpretation.
Many Seyfert galaxies, in particular, are amorphous or otherwise
disturbed (MacKenty 1989, 1990), and are morphologically similar
to the remnants formed by minor mergers with gas-rich disks.
While it is clearly premature to blame all, or even some
occurrences of Seyfert activity on minor mergers, we believe that
further investigation of this issue is warranted, based on the
results reported here.

\acknowledgments

This work was supported in part by the Pittsburgh Supercomputing
Center, the Alfred P. Sloan Foundation, NASA Grant NAGW--2422, and
the NSF under Grants AST 90--18526, ASC 93--18185 and the
Presidential Faculty Fellows Program.

\clearpage

\begin{figure}
\caption{Evolution of gas in isolated disk/halo system. Time is noted in
the upper right corner of each panel and individual frames
measure 20 length units per edge.}
\end{figure}

\begin{figure}
\caption{Evolution of the stellar disk in fiducial satellite merger,
seen face--on to the disk plane. In each panel, the cross locates
the instantaneous position of the satellite, and time is noted in
the upper right corner. The frames measure 20 length units per
edge.}
\end{figure}

\begin{figure}
\caption{Same as Figure 2, but seen edge--on to the disk plane.}
\end{figure}

\begin{figure}
\caption{Same as Figure 2, but showing the gas response in the disk.}
\end{figure}

\begin{figure}
\caption{Evolution of the satellite galaxy in fiducial satellite merger, seen
face--on to the disk plane.  Time is noted in the upper right corner
of each panel.  The frames measure 20 length units per edge.}
\end{figure}

\begin{figure}
\caption{Detailed view of the evolution of the gas in fiducial satellite
merger, seen face--on to the disk plane.  Crosses in each panel indicate the
location of the satellite, and time is noted in the upper right corner.
The frames measure 2 length units per edge.}
\end{figure}

\begin{figure}
\caption{Cumulative distribution of gas mass in the fiducial satellite merger.
Note that accretion of gas mass onto the central clump has effectively
ceased by T=81.6.}
\end{figure}

\begin{figure}
\caption{Evolution of orbital (L), spin (S), and total (J) angular momentum
of galaxy components in the fiducial satellite merger: a) halo,
b) disk stars, c) gas clump, d) satellite core.}
\end{figure}

\begin{figure}
\caption{Evolution of the torques experienced by gas in nuclear accumulation.
Panel (a): comparison between total torque and loss of spin angular
momentum; panel (b): decomposition of total torque into gravitational
and hydrodynamical components; panel (c) decomposition of gravitational
torque by mass component.  ``Disk'' refers to disk stars, ``halo'' to
halo particles, ``satellite'' to satellite particles, and ``gas'' to
the gas {\it not} driven into the central gas clump; panel (d):
decomposition of torque from disk stars into contribution from different
mass zones (see text).}
\end{figure}

\begin{figure}
\caption{Decomposition of stellar half-mass distribution into Fourier modes
during the merger.}
\end{figure}

\begin{figure}
\caption{Evolution of the gas in satellite merger with T$_{\rm gas} =
3\times 10^3$ K, seen face--on to the disk plane. The cross
locates the position of the satellite in each frame, and time is
indicated in the upper right corner. Each panel measures 2 length
units per edge.}
\end{figure}

\begin{figure}
\caption{Same as Figure 11, but for the simulation employing a T$_{\rm gas} =
10^5$ K gas.}
\end{figure}

\begin{figure}
\caption{Final cumulative gas mass distribution for various gas temperatures.}
\end{figure}

\begin{figure}
\caption{Distribution of gas densities following the merger in models
employing gases at different (isothermal) temperatures.  From
right to left, top to bottom, the frames show models with gas
temperatures $T=3\times 10^3, 10^4, 3\times 10^4, 10^5,$ and
$3\times 10^5$ K.}
\end{figure}

\begin{figure}
\caption{Evolution of the stellar disk in satellite merger with primary
which includes a dense bulge, seen face--on to the disk plane. In
each panel, the cross locates the instantaneous position of the
satellite, and time is noted in the upper right corner. The frames
measure 20 length units per edge.}
\end{figure}

\begin{figure}
\caption{Same as Figure 15, but showing the gas response in the disk.}
\end{figure}

\clearpage

\begin{figure}
\caption{Decomposition of stellar half-mass distribution into Fourier modes
during the merger involving the primary containing a dense central
bulge.}
\end{figure}

\begin{figure}
\caption{Detailed view of the evolution of the gas in satellite merger
with primary
which includes a dense bulge, seen face--on to the disk plane.  Crosses in
each panel indicate the location of the satellite, and time is noted in the
upper right corner.  The frames measure 2 length units per edge.}
\end{figure}

\begin{figure}
\caption{Final cumulative distribution of gas mass in the fiducial satellite
merger (solid line) and runs which include dense bulges of mass 1/3
(dash-dotted line) and 1/9 (dotted line). For comparison, the initial
gas mass distribution is also shown (dashed line).}
\end{figure}

\end{document}